\documentclass[letterpaper]{article} % DO NOT CHANGE THIS
\usepackage{aaai25}  % DO NOT CHANGE THIS
\usepackage{times}  % DO NOT CHANGE THIS
\usepackage{helvet}  % DO NOT CHANGE THIS
\usepackage{courier}  % DO NOT CHANGE THIS
\usepackage[hyphens]{url}  % DO NOT CHANGE THIS
\usepackage{graphicx} % DO NOT CHANGE THIS
\usepackage{natbib}  % DO NOT CHANGE THIS AND DO NOT ADD ANY OPTIONS TO IT
\usepackage{caption} % DO NOT CHANGE THIS AND DO NOT ADD ANY OPTIONS TO IT
\usepackage{algorithm}
\usepackage{algorithmic}
\usepackage{amsmath}
\usepackage{amssymb}
\usepackage{booktabs}
\usepackage{siunitx}
\usepackage{multirow}

\usepackage{newfloat}
\usepackage{listings}
\DeclareCaptionStyle{ruled}{labelfont=normalfont,labelsep=colon,strut=off} % DO NOT CHANGE THIS
\floatstyle{ruled}
\newfloat{listing}{tb}{lst}{}
\floatname{listing}{Listing}
\title{Region-Based Optimization in Continual Learning for Audio Deepfake Detection}
\author{
    Yujie Chen \textsuperscript{\rm 1},
    Jiangyan Yi \textsuperscript{\rm 3,*},
    Cunhang Fan \textsuperscript{\rm 1},
    Jianhua Tao \textsuperscript{\rm 3, 4},
    Yong Ren \textsuperscript{\rm 2},
    Siding Zeng \textsuperscript{\rm 2},
    Chu Yuan Zhang \textsuperscript{\rm 3},
    Xinrui Yan \textsuperscript{\rm 2},
    Hao Gu \textsuperscript{\rm 2},
    Jun Xue \textsuperscript{\rm 1},
    Chenglong Wang \textsuperscript{\rm 2}, 
    Zhao Lv \textsuperscript{\rm 1},
    Xiaohui Zhang \textsuperscript{\rm 2,\thanks{Corresponding authors.}}
}
\affiliations{
     \textsuperscript{\rm 1} School of Computer Science and Technology, Anhui University, China \\
    \textsuperscript{\rm 2} Institute of Automation, Chinese Academy of Sciences \\
    \textsuperscript{\rm 3} Department of Automation, Tsinghua University \\
    \textsuperscript{\rm 4} Beijing National Research Center for lnformation Science and Technology, Tsinghua University \\
    e22201148@stu.ahu.edu.cn, yijy@tsinghua.edu.cn
}

\usepackage{bibentry}
\begin{document}

\maketitle

\begin{abstract}
Rapid advancements in speech synthesis and voice conversion bring convenience but also new security risks, creating an urgent need for effective audio deepfake detection. Although current models perform well, their effectiveness diminishes when confronted with the diverse and evolving nature of real-world deepfakes. To address this issue, we propose a continual learning method named Region-Based Optimization (RegO) for audio deepfake detection. Specifically, we use the Fisher information matrix to measure important neuron regions for real and fake audio detection, dividing them into four regions. First, we directly fine-tune the less important regions to quickly adapt to new tasks. Next, we apply gradient optimization in parallel for regions important only to real audio detection, and in orthogonal directions for regions important only to fake audio detection. For regions that are important to both, we use sample proportion-based adaptive gradient optimization. This region-adaptive optimization ensures an appropriate trade-off between memory stability and learning plasticity. Additionally, to address the increase of redundant neurons from old tasks, we further introduce the Ebbinghaus forgetting mechanism to release them, thereby promoting the model’s ability to learn more generalized discriminative features. Experimental results show our method achieves a 21.3\% improvement in EER over the state-of-the-art continual learning approach RWM for audio deepfake detection. Moreover, the effectiveness of RegO extends beyond the audio deepfake detection domain, showing potential significance in other tasks, such as image recognition. The code is available at https://github.com/cyjie429/RegO
\end{abstract}


\section{Introduction}
Recently, with the rapid development of speech synthesis and voice conversion technologies, the distinction between real and fake audio has become increasingly blurred, posing significant security risks to society. Consequently, researchers are increasingly focusing on audio deepfake detection mechanisms \cite{yi2023audio}. Community-led initiatives, such as the ASVspoof Challenges \cite{kinnunen2017asvspoof, todisco2019asvspoof, yamagishi2021asvspoof} and the Audio Deepfake Detection (ADD) Challenges \cite{yi2022add, yi2023add}, significantly advance the field of fake audio detection. In addition, the introduction of pre-trained audio models significantly improves the effectiveness of audio deepfake detection, achieving impressive performance on publicly available datasets. \cite{tak2022automatic, wang22_odyssey, hsu2021hubert}

Despite the significant advancements in audio deepfake detection models, their performance is still limited when confronted with diverse and unseen forged audio in real-world scenarios. To address this challenge, two primary approaches have been developed. The first approach utilizes data augmentation and multi-feature fusion techniques to extract robust audio features, improving the generalization of model across various datasets \cite{wang23x_interspeech, fan2024frequency}. The second approach is based on continual learning \cite{ma21b_interspeech}, where the model incrementally learns from both new and old datasets, allowing it to integrate previously learned discriminative information. This enhances its detection capability for diverse and unseen deepfakes, achieving a balance between \textbf{memory stability} (the model's ability to retain performance on old tasks) and \textbf{learning plasticity} (the model's ability to perform on new tasks). Currently, the most advanced continual learning methods for audio deepfake detection, Regularized Adaptive Weight Modification (RAWM) \cite{zhang2023you} and Radian Weight Modification (RWM) \cite{zhang2024remember}, overcome catastrophic forgetting by introducing trainable gradient correction directions to optimize weights. While RAWM and RWM exhibit notable effectiveness in overcoming catastrophic forgetting, the use of Recursive Least Squares (RLS) \cite{shah1992optimal} to approximate the gradient plane of previous tasks can introduce errors \cite{738242}. Moreover, applying gradient modification to all neurons restricts the model's learning plasticity for new tasks.

To address the aforementioned issues, we propose a continual learning method for audio deepfake detection, named Region-Based Optimization (RegO). Under the same acoustic environments, real audio exhibits a more compact feature distribution compared to fake audio \cite{ma21b_interspeech, zhang2023you, zhang2024remember}, so they can be seen as a whole from the same dataset. Based on this observation, we propose that for new tasks, gradient updates for real audio should be parallel to the gradient directions of previous tasks, while the updates for fake audio should be orthogonal to the previous task gradients. Specifically, we utilize the Fisher information matrix (FIM) to measure the importance of neurons in the model, and then calculate the FIM for both real and fake audio detection separately. Following the fundamental principle of not constraining unimportant neurons to allow the model to quickly adapt to new tasks, and optimizing important neurons to overcome catastrophic forgetting, we divide the neurons into four regions for fine-grained region-adaptive gradient optimization: neurons that are unimportant for both real and fake audio detection are finetune directly to quickly adapt to new tasks; neurons that are important only for real audio detection are updated in parallel to the previous task gradients; neurons important only for fake audio detection are updated orthogonally to the previous task gradients; and neurons important for both real and fake audio detection are optimized through adaptive gradient updates based on the ratio of real to fake samples to maintain memory stability.

However, when the number of neurons in the model remains fixed as tasks increase, two problems arise: First, the number of neurons in less important regions decreases, making it progressively harder for the model to adapt to new tasks. Second, redundant neurons appear—those that are beneficial for only a few specific tasks but ineffective for others. To address these challenges, we draw inspiration from the non-linear nature of human memory forgetting \cite{loftus1985evaluating, ebbinghaus2013memory}. Over time, the memories that persist are generally those involving deeply understood knowledge, while other redundant Knowledge are forgotten. Based on this principle, we further propose a neuron forgetting mechanism inspired by the Ebbinghaus forgetting curve \cite{wozniak1995two} to release redundant neurons from previous tasks. This mechanism enables the model to learn knowledge from new tasks more efficiently, ensuring quicker adaptation to new tasks and the acquisition of more generalized discriminative information.

Our experiments on the Evolving Deepfake Audio (EVDA) benchmark \cite{zhang2024evda} demonstrate that our method outperforms several mainstream continual learning methods and state-of-the-art continual audio deepfake detection methods, including RAWM and RWM, in terms of balancing stability and plasticity. Furthermore, our method can be easily extended to other domains. General study conducted on image recognition tasks also showed competitive results, highlighting the potential significance of our approach across different machine learning fields. 

In summary, we make the following contributions:
\begin{itemize}

\item We propose a continual learning method for audio deepfake detection, named RegO, which partitions the neural network parameter space into four regions using the FIM. This method facilitates fine-grained, region-adaptive gradient optimization, ensuring an optimal trade-off between memory stability and learning plasticity.

\item We further propose a neuron forgetting mechanism based on Ebbinghaus forgetting curve, which releases redundant neurons from previous tasks to ensure faster adaptation to new tasks and to learn more generalizable discriminative information.

\item We conduct extensive experiments on the EVDA benchmark to validate the effectiveness of our method. Furthermore, we perform general study, and the results indicate that our approach holds potential significance in other domains, such as image recognition, without being limited to a specific field.
\end{itemize}

\section{Related Work}

Continual learning is a machine learning paradigm that aims to enable models to retain and use previously learned knowledge while continuously learning new tasks, thereby overcoming catastrophic forgetting. Current mainstream methods can be categorized into the following types: Regularization-based methods, which balance new and old tasks by selectively adding regularization terms to the changes in network parameters, such as Elastic Weight Consolidation (EWC) \cite{kirkpatrick2017overcoming}, Synaptic Intelligence (SI) \cite{zenke2017continual}, Gradient Episodic Memory (GEM) \cite{lopez2017gradient}, Orthogonal Weight Modification (OWM) \cite{zeng2019continual} etc. \cite{qiao2024prompt, elsayed2024addressing} Replay-based methods, which carefully select training samples from previous tasks, store them in a buffer, and mix them with new task training samples to ensure accuracy for old tasks, such as Greedy Sampler and Dumb Learner (GDumb) \cite{prabhu2020gdumb} and CWRStar \cite{lomonaco2020rehearsal}; Architecture-based methods, which use parameter isolation and dynamic expansion of the parameter space to protect previously acquired knowledge. \cite{wang2022coscl, razdaibiedina2023progressive}

Mainstream continual learning methods have achieved significant success in various fields, including image classification \cite{ razdaibiedina2023progressive, yoo2024layerwise}, object detection \cite{MENEZES2023476}, and semantic segmentation \cite{10208892, 10205421}. Continual learning methods are also effective in audio deepfake detection, empowering models to recognize diverse and unseen fake audio by leveraging continual learning principles. Currently, most continual learning methods for audio deepfake detection are regularization-based, such as Detecting Fake Without Forgetting (DFWF) \cite{ma21b_interspeech}, RAWM \cite{zhang2023you}, and RWM \cite{zhang2024remember},  and have demonstrated impressive results. However, the approximations made by regularization methods can lead to cumulative errors during continual learning, affecting the balance between model stability and plasticity. In contrast, our method selectively adjusts the gradients of parameters that are crucial for previous tasks while fine-tuning the remaining parameters to learn new tasks. This allows our method to adapt more quickly to new tasks while preventing catastrophic forgetting of old ones. 

\begin{figure*}[t]
    \centering
    \includegraphics[width=0.81\textwidth]{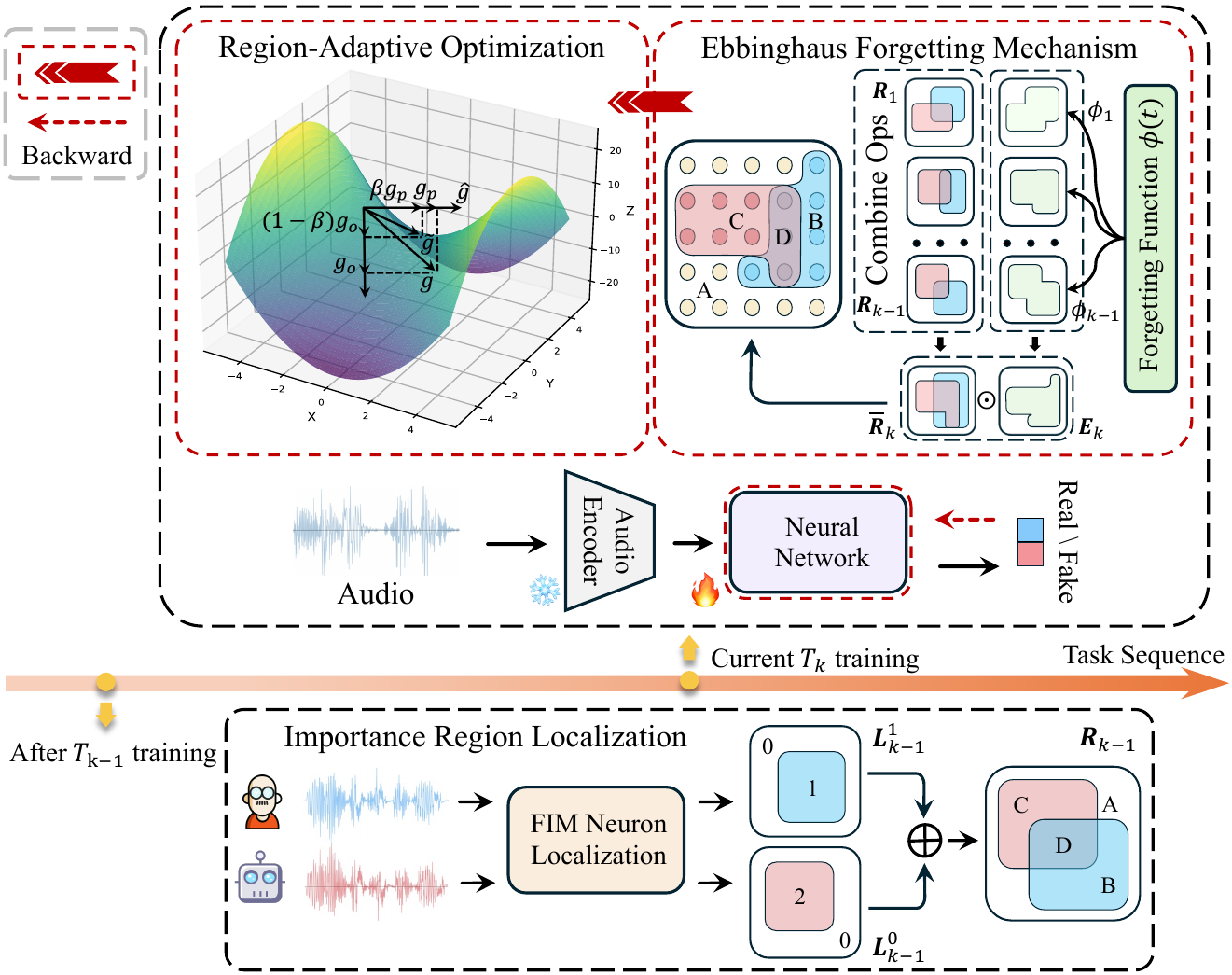}
    \caption{Illustration of RegO Architecture. (i).After the training of each task, we calculate the region matrix $\mathbf{R}$ through the IRL module. (ii).From the second task onward, gradient optimization is performed during backpropagation (shown by dark red arrows and boxes). The $\mathbf{E}_k$ is obtained through the EFM module and then combined with the historical $\mathbf{R}$ to generate $\overline{\mathbf{R}}$. (iii).Based on $\overline{\mathbf{R}}$, the RAO module updates the model weights as follows: Region A: fine-tuning (i.e. $g$); Region B: gradient update in the projection direction (i.e. $g_p$); Region C: gradient update in the orthogonal direction (i.e. $g_o$); Region D: adaptive gradient update based on the number of samples (i.e. $\widetilde{g}$).}
    \label{fig1:framework}
\end{figure*}

\section{Methodology}
In this section, the Region-Based Optimization (RegO) architecture, as illustrated in Figure 1, encompasses the principles and implementation of three core modules: Importance Region Localization (IRL), Region-Adaptive Optimization (RAO), and the Ebbinghaus Forgetting Mechanism (EFM).

\subsection{Definitions and Notation}
In Continual Learning, we define a sequence of tasks \(\{T_1, T_2, T_3, \ldots, T_N\}\) of \(N\) tasks. For the \(k\)-th task \(T_k\), there is a corresponding training set $\mathcal{D}_k = \left\{ \left( x_k^i, y_k^i \right) \right\}_{i=1}^{N_k} $ and corresponding parameters \(\theta_k\). The prediction of the model on input \( x \) is denoted by $f(x; \theta_k)$. During the training of task \( k \), we define the loss function as follows:
\begin{equation}
\mathcal{L}(\theta_k, \mathcal{D}_k) = \frac{1}{|\mathcal{D}_k|} \sum_{(x, y) \in \mathcal{D}_k} CE(f(x; \theta_k), y)
\end{equation}
where $CE$ is the standard cross-entropy loss. The gradient is shown as follows:
\begin{equation}
g_k = \nabla_{\theta_k} \mathcal{L}(\theta_k, D_k)
\end{equation}

\subsection{Importance Region Localization}
First, we need to identify the neuron regions that are important for the previous tasks. Inspired by \cite{kirkpatrick2017overcoming, Husz_r_2018}, we choose the Fisher Information Matrix (FIM) as a measure of neuron importance. After completing the training of the $k$-th task, we pass real and fake audio through the model separately to calculate the corresponding FIM, which encapsulate the importance measures of the weights for both real and fake audio detection. The FIM is defined as follows:
\begin{equation}
\mathbf{F}_k^{(c)} = \mathbb{E} \left[ \left. \nabla_{\theta} \log p(D_k^{(c)}|\theta) \nabla_{\theta} \log p(D_k^{(c)}|\theta)^\top \right|_{\theta = \theta_k^*} \right]
\end{equation}
Here, \(c\) is 0 (fake) or 1 (real), \(D_k\) represents the training dataset corresponding to $T_k$, and \(\theta_k^*\) denotes the optimal parameters for $T_k$. Note that the log probability of the data $D_k^{(c)}$ given the parameters $\log p(D_k^{(c)}|\theta)$ is simply the negative of the loss function $-\mathcal{L}(\theta_k, \mathcal{D}_k)$ for task $T_k$. Based on this, by setting the threshold \(\alpha\), we locate important neurons, resulting in a localization matrix, as shown in Equation 4. 
\begin{equation}
\mathbf{L}_k^{(c)}[i][j] =
\begin{cases}
2 & \text{if } c = 0 \text{ and } \mathbf{F}_k^{(c)}[i][j] \geq \mathrm{P}_{\alpha}(\mathbf{F}_k^{(c)}) \\
1 & \text{if } c = 1 \text{ and } \mathbf{F}_k^{(c)}[i][j] \geq \mathrm{P}_{\alpha}(\mathbf{F}_k^{(c)}) \\
0 & \text{otherwise}
\end{cases}
\end{equation}
where \(i, j\) represents the neuron index, \( P_\alpha \) represents the \(\alpha\)-percentile. Based on this, by summing the localization matrices for both real and fake audio, we can identify four distinct regions, as shown in Equation 5.
\begin{equation}
\mathbf{R}_k[i][j] = \mathbf{L}_k^{\text{0}}[i][j] + \mathbf{L}_k^{\text{1}}[i][j]
\end{equation}
In the \(\mathbf{R}_k\), regions with values 0, 1, 2, and 3 are denoted by the letters A, B, C, and D, respectively. Region A represents neurons that are not important for $T_k$. Region B represents neurons that are important for real audio detection in $T_k$. Region C represents neurons that are important for fake audio detection in $T_k$. Region D represents neurons that are important for both real and fake audio detection in $T_k$.

\subsection{Region-Adaptive Optimization}
During training, starting from the second task, we merge all region matrices $\mathbf{R}$ from the previous tasks to obtain $\overline{\mathbf{R}}$. For the four regions $A$, $B$, $C$, and $D$ within $\overline{\mathbf{R}}$, we adaptively optimize each region according to the following principles.

Firstly, because the neurons in region A have minimal impact on previous tasks, but are likely to become important for new tasks, we allow them to quickly adapt and learn the knowledge of the new tasks. Therefore, we do not apply additional gradient optimization to the neurons in region A and instead update them using fine-tuning, the gradient of region A is defined in Equation 6.
\begin{equation}
    g_A = g \odot \mathbb{I}_{\{\overline{\mathbf{R}}[i][j] = 0\}}
\end{equation}
where \(\odot\) denote the Hadamard product, $\mathbb{I}_{\{\overline{\mathbf{R}}[i][j]=0\}}$ is an indicator function that takes a value of 1 when \(\overline{\mathbf{R}}[i][j]\) equals 0 and 0 otherwise. 

Second, as mentioned above, real audio has a more compact feature distribution compared to fake audio and can be considered as coming from the same dataset. Therefore, to largely retain the knowledge of real audio detection from previous tasks, the neurons in region B should project the current gradient $g$ onto the direction of the old task gradient $\hat{g}$ for gradient optimization, as shown in Equation 7.
\begin{equation}
\begin{aligned}
    g_p &= \frac{g \cdot \hat{g}}{\| \hat{g} \|^2} \hat{g} \\
    g_B &= g_p \odot \mathbb{I}_{\{\overline{\mathbf{R}}[i][j] = 1\}}
\end{aligned}
\end{equation}

\begin{algorithm}[t]
\caption{Region-Based Optimization}
\textbf{Require:} Training data from different datasets, $\eta$ (learning rate), $\mathcal{R}$ (Region matrix set)
\begin{algorithmic}[1]
\FOR{every dataset $k$}
\FOR{every batch $b$}
    \IF{k = 1}
        \STATE Update $\theta_k$: $\theta_k \leftarrow \theta_k - \eta w$
    \ELSE
        \STATE Compute memory matrix $\mathbf{M}_k$ by Equ.(13)
        \STATE Compute Ebbinghaus matrix $\mathbf{E}_k$ by Equ.(14)
        \STATE Compute $\overline{\mathbf{R}}_k$ by combining region matrix set $\mathcal{R}$
        \STATE $ g_A \leftarrow g \odot \mathbb{I}_{\{\overline{\mathbf{R}}_k[i][j] = 0\}} $
        \STATE $ g_p \leftarrow \frac{g \cdot \hat{g}}{\| \hat{g} \|^2} \hat{g} $
        \STATE $ g_B \leftarrow g_p \odot \mathbb{I}_{\{\overline{\mathbf{R}}_k[i][j] = 1\}} $
        \STATE $ g_o \leftarrow g - g_p $
        \STATE $ g_C \leftarrow g_o \odot \mathbb{I}_{\{\overline{\mathbf{R}}_k[i][j] = 2\}} $
        \STATE $ \beta \leftarrow \frac{\sum_{l=1}^{u} N^l}{\sum_{l=1}^{u+v} N^l} $
        \STATE $ \widetilde{g} \leftarrow \beta * g_p + (1-\beta) * g_o $
        \STATE $ g_D \leftarrow \widetilde{g} \odot \mathbb{I}_{\{\overline{\mathbf{R}}_k[i][j] = 3\}} $
        \STATE Initialization: $ w \leftarrow 0 $ 
        \STATE $ w \leftarrow g_A + g_B + g_C + g_D $
        \STATE Update $\theta_k$: $\theta_k \leftarrow \theta_k - \eta w$
    \ENDIF    
\ENDFOR
\STATE Compute the k-th Region Matrix $\mathbf{R}_k$ by Equ.(1)(2)(3)
\STATE $\mathcal{R} \leftarrow \mathbf{R}_k$
\ENDFOR
\end{algorithmic}
\end{algorithm}

Thirdly, due to the diversity of speech synthesis and voice conversion methods, there is a wide variance in feature distributions among fake audio samples. Therefore, to reduce the forgetting of discriminative information about fake audio from previous tasks while learning discriminative information for fake audio in new tasks, we update the gradient direction of neurons in region C to be orthogonal to the gradient direction of the old tasks, as shown in Equation 8.
\begin{equation}
\begin{aligned}
    g_o &= g - g_p \\
    g_C &= g_o \odot \mathbb{I}_{\{\overline{\mathbf{R}}[i][j] = 2\}}
\end{aligned}
\end{equation}

Fourth, region D is crucial for both real and fake audio discrimination from previous tasks. Therefore, to balance the retention of knowledge for both real and fake audio detection in neurons of region D, we need to optimize the gradient update direction to achieve an optimal trade-off. Specifically, we adaptively determine whether the gradient update direction should lean more towards the projection direction or the orthogonal direction based on the proportion of real and fake audio samples, as shown in Equations 9 and 10.
\begin{equation}
\beta = \frac{\sum_{l=1}^{u} N^l}{\sum_{l=1}^{u+v} N^l}
\end{equation}
\begin{equation}
\begin{aligned}
    \widetilde{g} = \beta * g_p + (1-\beta) * g_o \\
    g_D = \widetilde{g} \odot \mathbb{I}_{\{\overline{\mathbf{R}}[i][j] = 3\}}
\end{aligned}
\end{equation}
In Equation 9, \(u\) and \(v\) represent the number of classes with similar feature distributions and the remaining classes, respectively. In deepfake audio detection, \(u\) and \(v\) are both set to 1, indicating the two classes of real and fake audio. In image recognition, \(u\) and \(v\) represent the number of classes with similar feature distributions and the number of classes with dissimilar feature distributions, respectively. \(N^l\) denotes the number of samples in the batch for the \(l\)-th class.

Finally, the total gradient update for a batch is defined as:
\begin{equation}
w = g_A + g_B + g_C + g_D
\end{equation}

\begin{table*}[t]
\centering
\begin{tabular}{lccccccccc}
\toprule
\multicolumn{1}{c}{\multirow{2}{*}{Continual Learning Methods}} & \multicolumn{9}{c}{EER ($\downarrow$) on each experience}\\  
\cmidrule(r){2-10}
\multicolumn{1}{c}{} & $\rm Exp_1$  & $\rm Exp_2$ & $\rm Exp_3$ & $\rm Exp_4$ & $\rm Exp_5$ & $\rm Exp_6$ & $\rm Exp_7$ & $\rm Exp_8$ & Avg\\ 
\midrule
\multicolumn{1}{c}{Replay-All} & 2.80 & 5.68 & 1.52 & 0.76 & 1.84 & 7.96 & 5.76 & 2.56 & 3.61\\  
\multicolumn{1}{c}{Finetune-$\rm Exp_{1}$} & 2.20 & 24.80 & 23.16 & 16.84 & 23.80 & 34.12 & 26.44 & 15.52 & 20.86\\ 
\midrule
\multicolumn{1}{c}{Finetune} & 5.16 & 15.56 & 8.20 & 2.32 & 4.08 & 21.72 & 9.64 & \underline{3.04} & 8.72\\ 
\multicolumn{1}{c}{EWC} & \textbf{3.72} & 13.92 & 7.32 & 2.12 & \underline{3.56} & \underline{17.40} & 10.24 & 3.16 & \underline{7.68} \\
\multicolumn{1}{c}{GDumb} & 4.72 & 14.12 & 7.32 & 4.60 & 6.56 & 24.28 & 15.28 & 11.40 & 11.03 \\
\multicolumn{1}{c}{GEM} & 5.60 & 16.56 & 6.28 & 2.60 & 9.60 & 24.44 & 11.88 & 4.28 & 10.15 \\
\multicolumn{1}{c}{CWRStar} & 5.12 & 27.92 & 22.88 & 29.36 & 45.52 & 43.20 & 49.92 & 18.32 & 30.28 \\
\multicolumn{1}{c}{SI} & 6.96 & \underline{10.88} & \underline{5.92} & \underline{1.60} & 4.04 & 18.96 & 10.04 & 3.32 & 7.71 \\
\multicolumn{1}{c}{OWM} & 27.28 & 33.72 & 29.32 & 33.12 & 47.28 & 49.52 & 48.80 & 26.32 & 36.92 \\
\multicolumn{1}{c}{RAWM} & 9.28 & 16.04 & 6.76 & 2.60 & 3.60 & 19.52 & \underline{9.64} & 3.40 & 8.85 \\
\multicolumn{1}{c}{RWM} & 4.44 & 14.92 & 6.28 & 1.92 & 4.44 & 18.92 & 10.04 & 3.52 & 8.06 \\
\midrule \midrule
\multicolumn{1}{c}{RegO (Ours)} & \underline{4.36} & \textbf{10.64} & \textbf{3.76} & \textbf{1.20} & \textbf{3.16} & \textbf{15.72} & \textbf{9.16} & \textbf{2.72} & \textbf{6.34} \\
\bottomrule
\end{tabular}
\caption{The EER (\%) of our method compared with various methods.}
\label{table 1}
\end{table*}

\subsection{Ebbinghaus Forgetting Mechanism}
During the continual learning process, when the number of neurons remains constant, neurons in region A gradually diminish, and redundant neurons that only benefit individual tasks begin to emerge, which undermines the model's adaptability and generalization ability. To address this issue, inspired by Ebbinghaus forgetting theory \cite{loftus1985evaluating, ebbinghaus2013memory}, we propose a neuron forgetting mechanism based on the Ebbinghaus memory curve \cite{wozniak1995two}. The approximation function is defined as Equation 12.
\begin{equation}
\phi(t) = e^{-\frac{t}{k}}
\end{equation}
where $t$ represents the time steps and $k$ denote the $k$-th task. Specifically, we define the Ebbinghaus forgetting curve function based on the number of tasks processed so far, and calculate the memory weights using this forgetting curve function. Then, we allocate memory weights to the region matrix $\mathbf{R}$ of the old tasks. By Equation 13 and 14, we compute the memory matrix $\mathbf{M}$, which contains the accumulated memory weights for each neuron. Finally, we set a threshold $\gamma$. When the memory weight of a neuron is less than $\gamma$, that neuron is released, resulting in the Ebbinghaus matrix $\mathbf{E}$.
\begin{equation}
\mathbf{M}_k[i][j] = \sum_{t=1}^{k-1} \phi(t) \ast \mathbb{I}_{\{\mathbf{R}_t[i][j] \in \{0,1,2,3\}\}}
\end{equation}
\begin{equation}
\mathbf{E}_k[i][j] = 
\begin{cases} 
1 & \text{if } \mathbf{M}_k[i][j] > \gamma \\
0 & \text{if } \mathbf{M}_k[i][j] \leq \gamma \\
\end{cases}
\end{equation}

\section{Experiments}

We conduct a series of experiments to evaluate the effectiveness of our approach. The experiments are performed on a continual learning benchmark EVDA \cite{zhang2024evda} for speech deepfake detection, which includes eight publicly available and popular datasets specifically designed for incremental synthesis algorithm audio deepfake detection. Additionally, we carry out a general study in the field of image recognition using the well-established continual learning benchmark, CLEAR \cite{lin2021clear}.

\begin{table*}[t]
\centering
\begin{tabular}{lccccccccc}
\toprule
\multicolumn{1}{c}{\multirow{2}{*}{Ablation Study}} & \multicolumn{9}{c}{EER ($\downarrow$) on each experience}\\  
\cmidrule(r){2-10}
\multicolumn{1}{c}{} & $\rm Exp_1$  & $\rm Exp_2$ & $\rm Exp_3$ & $\rm Exp_4$ & $\rm Exp_5$ & $\rm Exp_6$ & $\rm Exp_7$ & $\rm Exp_8$ & Avg\\ 
\midrule
\multicolumn{1}{c}{RegO (Ours)} & \textbf{4.36} & 10.64 & \textbf{3.76} & \textbf{1.20} & 3.16 & \textbf{15.72} & \textbf{9.16} & \textbf{2.72} & \textbf{6.34} \\
\multicolumn{1}{c}{w/o EFM} & 4.48 & \textbf{10.48} & 5.56 & 1.32 & \textbf{3.12} & 16.48 & 9.68 & 3.32 & 6.80 \\
\multicolumn{1}{c}{w/o IRL} & 5.32 & 12.28 & 4.72 & 1.52 & 3.48 & 19.08 & 9.76 & 2.92 & 7.38 \\
\multicolumn{1}{c}{w/o RAO} & 4.68 & 14.68 & 5.52 & 2.48 & 4.44 & 16.16 & 9.80 & 3.32 & 7.63 \\
\bottomrule
\end{tabular}
\caption{The EER (\%) results of the ablation study for our method.}
\label{table 2}
\end{table*} 

\subsection{Experimental Setup}

\subsubsection{Datasets and Metrics}
In this paper, we refer to each dataset as “$\rm Exp$” (e.g., $\rm Exp_1$, $\rm Exp_2$, ..., $\rm Exp_{10}$), representing the different datasets used in our experiments. The EVDA benchmark from $\rm Exp_1$ to $\rm Exp_8$ are FMFCC \cite{zhang2021fmfcc}, In the Wild \cite{muller22_interspeech}, ADD 2022 \cite{yi2022add}, ASVspoof2015 \cite{wu2017asvspoof}, ASVspoof2019 \cite{todisco2019asvspoof}, ASVspoof2021 \cite{yamagishi2021asvspoof}, FoR \cite{reimao2019dataset}, and HAD \cite{yi21_interspeech}. For the EVDA baseline, 2000 samples are randomly sampled from each dataset as the training set, and 5000 samples are sampled as the test set. The EVDA baseline dataset configuration includes cross-lingual (Chinese and English) and cross-task (whole-segment and partial-segment fake detection) scenarios to simulate the unseen and diverse real-world forgery conditions. The final model in this study refers to the model trained sequentially on these eight datasets and evaluated on each dataset. We use the standard metric Equal Error Rate (EER) \cite{wu2017asvspoof} in the field of audio deepfake detection to evaluate the performance of our model.

\subsubsection{Model}
We employ the pre-trained speech model Wav2vec 2.0 \cite{baevski2020wav2vec} as the feature extractor, the parameters of Wav2vec 2.0 are loaded from the pretrained model XLSR-53 \cite{conneau21_interspeech}. Given the robustness of the speech features obtained from the pre-trained model, we opt for a 5-layer SimpleMlp as the backend, which consists of fully connected layers with the following dimensions: 1024 to 512, 512 to 512 (x3), and 512 to 2.

\subsubsection{Training Details}
We use the Adam optimizer to finetune the SimpleMlp, with a learning rate $\eta$ of 0.0001 and a batch size of 32, on an NVIDIA A100 80GB GPU. To evaluate the performance of our method, we compare it with six widely used continual learning methods, finetuning, and two advanced continual learning methods specifically designed for audio deepfake detection: RAWM \cite{zhang2023you}, RWM \cite{zhang2024remember}. Additionally, we present the training results on all datasets (Replay-All), which are considered the lower bound for all mentioned continual learning methods \cite{parisi2019continual}.

\subsection{Comparison with Other Methods}
In this experiment, we compare RegO with other methods. Here, Finetune-$\rm Exp_{1}$ shows the results of training on $\rm Exp_{1}$ and evaluating on the other Exps, highlighting the significant differences between the various Exps. As shown in Table 1, after training on 8 $\rm Exps$, our method achieves the best performance on 7 of the $\rm Exps$ and the second-best performance on Exp1. The overall evaluation metrics demonstrate that our method is only slightly inferior compared to Replay-ALL, which is considered the upper bound for continual learning performance. Additionally, among the eight $\rm Exps$, $\rm Exp_1$, $\rm Exp_3$, and $\rm Exp_8$ are Chinese datasets, while the remaining ones are English datasets. Notably, $\rm Exp_3$ consists of low-quality speech data, and $\rm Exp_8$ is a partial-fake spoofing dataset. The experimental results demonstrate that our method shows promising potential in both cross-lingual and cross-task scenarios, indicating its capability to handle relatively diverse real-world audio deepfake environments.

\subsection{Ablation Study}
We conduct an ablation study to evaluate the effectiveness of the proposed modules. The results, shown in Table \ref{table 3}, are as follows: “w/o EFM” denotes the removal of the Ebbinghaus forgetting mechanism, “w/o IRL” indicates no division between real and fake regions, and “w/o RAO” refers to applying orthogonal gradient optimization to all weights. The results for “w/o RAO” suggest that, compared to orthogonally optimizing all weights, applying region-adaptive gradient optimization to critical regions alone achieves a better balance between model memory stability and learning plasticity. Additionally, we observe that “w/o EFM” performs better than RegO on some of the earlier datasets, but its overall capability is inferior to RegO. We attribute this to the significant differences among the eight audio deepfake detection datasets, which result in a substantial number of redundant neurons. The EFM module effectively filters out these redundant neurons, enabling the model to adapt more quickly to other tasks.

\begin{figure}[t]
    \centering
    \includegraphics[width=\columnwidth]{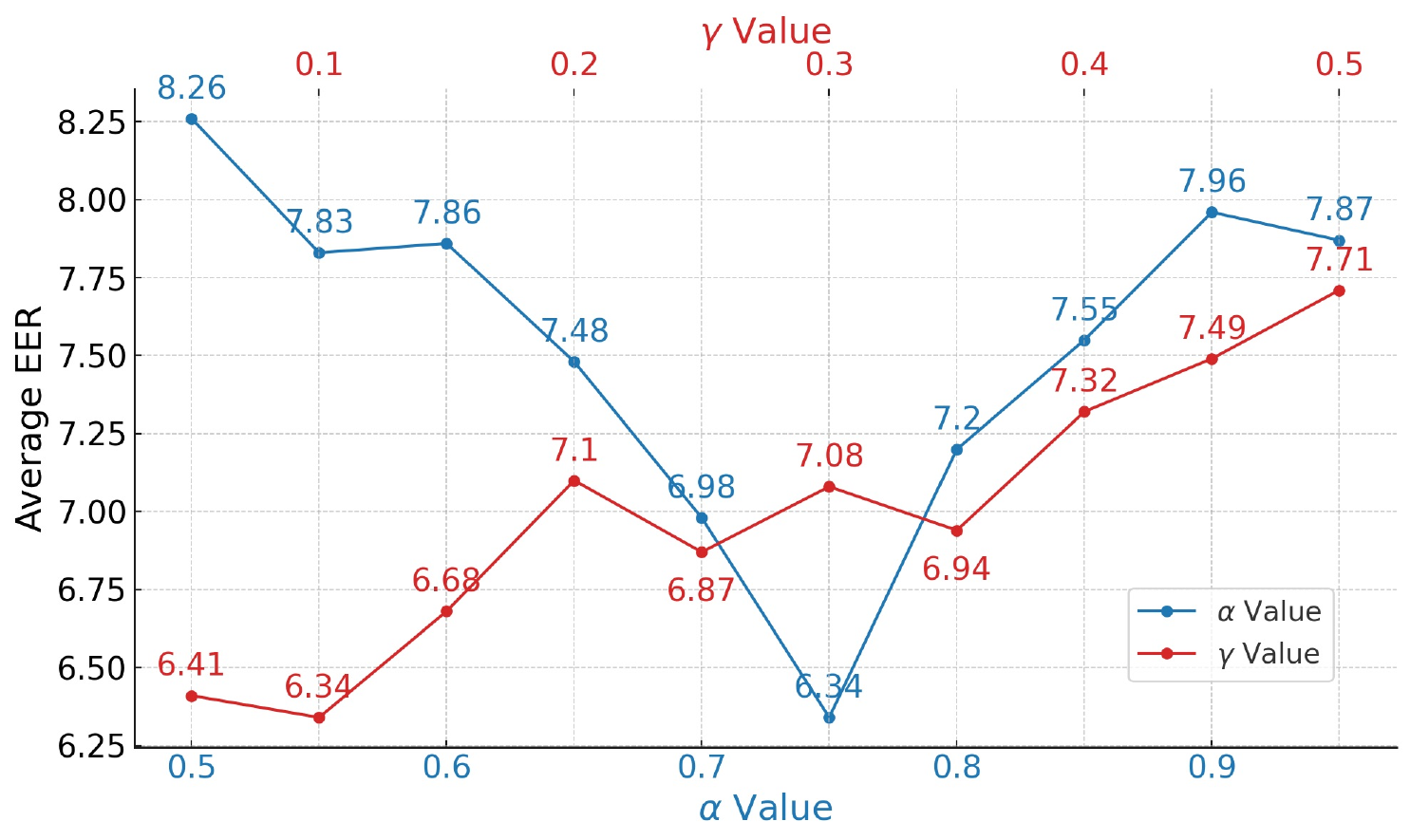}
    \caption{The average EER (\%) results of the hyperparameter study for our method RegO.}
    \label{fig2:Hyperparameter Study}
\end{figure}

\subsection{Hyperparameter Study}
We conduct a hyperparameter study to evaluate the impact of $\alpha$ and $\gamma$ on our method RegO. Notably, the $\alpha$ study is performed with $\gamma$ fixed at 0.1, while the $\gamma$ study is conducted with $\alpha$ fixed at 0.75. The experimental results show that our method performs best when $\alpha$ is set to 0.75. As $\alpha$ increases, the region of important neurons (i.e., regions B, C, and D) shrinks, reducing model stability. Conversely, as $\alpha$ decreases, the region of less important neurons (i.e., region A) diminishes, leading to reduced model plasticity. Both scenarios result in a decline in model performance. In the $\gamma$ experiments, increasing $\gamma$ causes more neurons to be classified as redundant, including important neurons effective across multiple early tasks that are mistakenly classified as redundant. This misclassification reduces model stability, leading to a decline in performance.

\begin{table*}[t]
\centering
\begin{tabular}{lcccccccccc}
\toprule
\multicolumn{1}{c}{\multirow{2}{*}{Continual Learning Methods}} & \multicolumn{10}{c}{Accuracy ($\uparrow$) on each experience}\\  
\cmidrule(r){2-11}
\multicolumn{1}{c}{} & $\rm Exp_1$  & $\rm Exp_2$ & $\rm Exp_3$ & $\rm Exp_4$ & $\rm Exp_5$ & $\rm Exp_6$ & $\rm Exp_7$ & $\rm Exp_8$ & $\rm Exp_9$ & $\rm Exp_{10}$\\ 
\midrule
\multicolumn{1}{c}{Replay-All} & 94.34 & 94.44 & 94.44 & 94.85 & 95.66 & 94.14 & 93.94 & 95.86 & 94.24 & 95.56 \\  
\multicolumn{1}{c}{Finetune-$\rm Exp_{1}$} & 57.27 & 56.67 & 59.60 & 58.89 & 59.39 & 55.05 & 56.16 & 54.75 & 54.65 & 55.15 \\ 
\midrule
\multicolumn{1}{c}{Finetune} & 90.40 & 89.80 & 90.10 & 92.73 & 90.71 & \underline{90.40} & 90.10 & 89.90 & 90.40 & 92.42 \\ 
\multicolumn{1}{c}{EWC} & \underline{91.72} & \underline{91.62} & \underline{91.31} & 92.12 & 91.31 & 90.40 & \underline{91.11} & \underline{90.61} & 90.71 & 93.13 \\ 
\multicolumn{1}{c}{GEM} & 91.62 & 90.51 & 90.30 & \underline{92.93} & \underline{91.62} & 89.39 & 90.30 & 90.10 & 89.49 & \underline{93.33} \\ 
\multicolumn{1}{c}{GDumb} & 90.20 & 87.78 & 89.60 & 89.09 & 89.19 & 86.57 & 87.47 & 88.18 & 87.88 & 88.38 \\ 
\multicolumn{1}{c}{CWRStar} & 87.68 & 87.98 & 87.58 & 88.79 & 88.79 & 86.77 & 87.58 & 86.77 & 86.77 & 90.00 \\ 
\multicolumn{1}{c}{SI} & 89.39 & 89.29 & 90.00 & 91.11 & 89.79 & 88.69 & 89.39 & 89.90 & 88.79 & 90.40 \\ 
\multicolumn{1}{c}{OWM} & 73.03 & 71.41 & 70.30 & 73.13 & 71.01 & 68.99 & 69.70 & 67.27 & 70.30 & 69.70 \\ 
\multicolumn{1}{c}{RAWM} & 85.25 & 84.95 & 82.83 & 83.84 & 84.14 & 81.62 & 81.52 & 83.64 & 83.84 & 82.42 \\ 
\multicolumn{1}{c}{RWM} & 87.17 & 86.26 & 87.68 & 89.29 & 87.17 & 85.66 & 88.18 & 85.15 & 86.87 & 85.86 \\ 
\midrule \midrule
\multicolumn{1}{c}{RegO (Ours)} & \textbf{91.92} & \textbf{93.03} & \textbf{92.63} & \textbf{93.64} & \textbf{93.94} & \textbf{92.32} & \textbf{92.53} & \textbf{92.53} & \textbf{92.42} & \textbf{94.75} \\
\bottomrule
\end{tabular}
\caption{The accuracy (\%) of the models trained on all CLEAR experiments. All results are reproduced by us.}
\label{table 3}
\end{table*}

\begin{table*}[t]
\centering
\begin{tabular}{lcccccccccc}
\toprule
\multicolumn{1}{c}{\multirow{2}{*}{Ablation Study}} & \multicolumn{10}{c}{Accuracy ($\uparrow$) on each experience}\\  
\cmidrule(r){2-11}
\multicolumn{1}{c}{} & $\rm Exp_1$  & $\rm Exp_2$ & $\rm Exp_3$ & $\rm Exp_4$ & $\rm Exp_5$ & $\rm Exp_6$ & $\rm Exp_7$ & $\rm Exp_8$ & $\rm Exp_9$ & $\rm Exp_{10}$\\ 
\midrule
\multicolumn{1}{c}{RegO (Ours)} & 91.92 & 93.03 & 92.63 & 93.64 & 93.94 & \textbf{92.32} & 92.53 & 92.53 & 92.42 & 94.75 \\
\multicolumn{1}{c}{w/o EFM} & \textbf{92.93} & \textbf{93.93} & \textbf{93.33} & \textbf{94.65} & \textbf{94.34} & 91.82 & \textbf{93.23} & \textbf{93.64} & \textbf{93.13} & \textbf{94.85} \\
\multicolumn{1}{c}{w/o IRL} & 90.61 & 89.70 & 90.61 & 91.21 & 91.41 & 88.89 & 90.71 & 89.49 & 89.49 & 92.83 \\
\multicolumn{1}{c}{w/o RAO} & 88.59 & 87.37 & 88.48 & 89.49 & 88.99 & 87.98 & 87.67 & 88.48 & 87.37 & 90.91 \\
\bottomrule
\end{tabular}
\caption{The accuracy (\%) results of the ablation study for our method on the CLEAR experiences.}
\label{table 4}
\end{table*}

\subsection{General Study}

\subsubsection{Experimental Setup}

We use the CLEAR benchmark \cite{lin2021clear} to evaluate the scalability of our method. CLEAR is a classic continual image classification benchmark, with datasets based on the natural temporal evolution of visual concepts in the real world. It adopts task-based sequential learning by dividing the temporal stream into 10 buckets, each composed of labeled subsets for training and evaluation (with 300 images in the training set and 150 in the test set), resulting in a series of 11-way classification tasks. A small labeled subset ($\rm Exp_1$, $\rm Exp_2$, ..., $\rm Exp_{10}$) consists of 11 temporally dynamic categories, including examples like computers, cosplay, etc. We use classification accuracy to evaluate model performance. For the image recognition model, we use a pre-trained ResNet-50 \cite{he2016deep} as the feature extractor, which is frozen during continual learning, generating 2048-dimensional features. The downstream classifier has three linear layers: 2048 to 1024, 1024 to 512, and 512 to 2. We set the initial learning rate to 0.1, a batch size of 512, and used SGD optimizer with 0.9 momentum.

\subsubsection{Comparison with Other Methods}
We compare RegO with several classic continual learning methods. As shown in Table \ref{table 3}, after training on 10 $\rm Exps$, the performance of RegO is second only to Replay-All, which is considered the upper bound for continual learning performance. Additionally, Table \ref{table 3} shows that RWM and RAWM perform better in the earlier subset ($\rm Exp_1$ - $\rm Exp_5$) compared to the later ones, indicating that these algorithms are more focused on mitigating forgetting, but are less adaptable to new tasks. Our method overcomes catastrophic forgetting by optimizing the gradients of important neurons, while fine-tuning less critical neurons directly to ensure rapid adaptation to new tasks, ensuring an appropriate stability-plasticity trade-off.

\subsubsection{Ablation Study}
We conduct an ablation study to evaluate the effectiveness of the proposed modules, with the results shown in Table \ref{table 4}. Compared to Table 3, we observe an interesting phenomenon: in the image recognition task, removing the EFM module leads to better model performance, which contrasts with the ablation results for audio deepfake detection. We speculate that this is because the CLEAR benchmark represents a decade-long natural temporal evolution of real-world visual concepts, where the appearance of major categories such as computers, cameras, etc. has remained relatively unchanged over the years. The results of Finetune-$\rm Exp_{1}$ support this hypothesis: after training on $\rm Exp_1$, the accuracy remains similar from $\rm Exp_{1}$ to $\rm Exp_{5}$ but shows a slight decline from $\rm Exp_{5}$ to $\rm Exp_{10}$. This indicates that retaining old knowledge might interfere with performance across multiple tasks. On the other hand, in the audio deepfake detection task, where differences in synthesis or conversion algorithms are more distinct (as shown by the Finetune-$\rm Exp_{1}$ results in Table 1), the role of EFM becomes more critical. Nevertheless, regardless of whether the EFM is integrated, both versions of our method consistently outperform other methods.

\section{Conclusion}
In this paper, we propose an effective continual learning algorithm, Region-Based Optimization (RegO), aimed at improving the generalization of audio deepfake detection models against diverse and unseen forgeries in real-world scenarios. The core idea of RegO is to avoid constraints on less important neurons, allowing the model to quickly adapt to new tasks, while applying region-based adaptive gradient optimization to important neurons to overcome catastrophic forgetting, achieving a suitable balance between memory stability and learning plasticity.  Experimental results demonstrate that our method outperforms SOTA method RWM for audio deepfake detection, proving its robustness against diverse forgery techniques. Additionally, we conduct the general study and achieve competitive results, indicating our method has potential significance in other domains, such as image recognition. Moreover, we plan to explore how to extend our method to address other challenges in machine learning, such as multi-task learning \cite{langa2021deepfakes}.

\section{Acknowledgments}
This work is supported by the National Natural Science Foundation of China (NSFC) (No.62322120, No.U21B2010, No.62306316, No.62206278, No.62201002, 6247077204), the {STI 2030—Major Projects (No. 2021ZD0201500)}, Excellent Youth Foundation of Anhui Scientific Committee (No. 2408085Y034), Distinguished Youth Foundation of Anhui Scientific Committee (No. 2208085J05), Special Fund for Key Program of Science and Technology of Anhui Province (No. 202203a07020008), Cloud Ginger XR-1.

\bibliography{aaai25}

\end{document}